\documentclass[aps,twocolumn,prx,amsfonts,showpacs,superscriptaddress,longbibliography]{revtex4-2}
\usepackage{graphicx}
\usepackage{bm}
\usepackage{hyperref}
\usepackage{amsmath}
\usepackage{amssymb}
\usepackage[table]{xcolor}
\usepackage{array}
\usepackage{multirow}
\usepackage[caption=false]{subfig}
\usepackage{physics}
\usepackage{booktabs}
\usepackage{soul}
\usepackage{amsthm}
\usepackage{bbm}
\usepackage{multirow}
\usepackage[colorinlistoftodos]{todonotes}
\usepackage[scr=boondoxo,frak=boondox,scrscaled=1.05]{mathalfa}
\usepackage{amsmath}

\def\r{{\boldsymbol{r}}}
\def\f{{\boldsymbol{f}}}

\def\k{{\boldsymbol{k}}}

\def\v{{\boldsymbol{v}}}
\def\q{{\boldsymbol{q}}}
\def\A{{\boldsymbol{A}}}
\def\g{{\boldsymbol{g}}}

\def\G{{\boldsymbol{G}}}

\def\b{{\boldsymbol{b}}}
\def\K{{\boldsymbol{K}}}
\def\L{{\boldsymbol{L}}}
\def\h{{\boldsymbol{h}}}
\def\w{{\boldsymbol{w}}}

\usepackage{ulem}
\begin{document}

\title{First-Principles AI finds crystallization of fractional quantum Hall liquids}

\author{Ahmed Abouelkomsan}
\thanks{ahmed95@mit.edu}
\affiliation{Department of Physics, Massachusetts Institute of Technology, Cambridge, MA-02139, USA}

\author{Liang Fu}
\thanks{liangfu@mit.edu}
\affiliation{Department of Physics, Massachusetts Institute of Technology, Cambridge, MA-02139, USA}

\begin{abstract}
When does a fractional quantum Hall (FQH) liquid crystallize? Addressing this question requires a framework that treats fractionalization and crystallization on equal footing, especially in strong Landau-level mixing regime. Here, we introduce MagNet, a self-attention neural-network variational wavefunction designed for quantum systems in magnetic fields on the torus geometry. We show that MagNet provides a \textit{unifying} and expressive ansatz capable of describing both FQH states and electron crystals within the same architecture. Trained solely by energy minimization of the microscopic Hamiltonian, MagNet discovers topological liquid and electron crystal ground states across a broad range of Landau-level mixing. Our results highlight the power of first-principles AI for solving strongly interacting many-body problems and finding competing phases without external training data or physics pre-knowledge. 

\end{abstract}
\maketitle
\date{\today}

\textit{Introduction} --- When does a fractional quantum Hall (FQH) liquid crystallize? This question lies at the heart of the competition between topological order and charge ordering in two-dimensional electron systems and has motivated numerous theoretical and numerical efforts \cite{ortiz1993new,zhao2018crystallization,price1993fractional,platzman1993quantum,zhu1993wigner,he2005phase}. It is also directly relevant experimentally \cite{santos1992observation,santos1992effect,maryenko2018composite,pan2005transition,tsui2024direct}: by tuning the carrier density and magnetic field, experiments can access both FQH and Wigner crystal regimes in high-mobility two dimensional semiconductors \cite{pack2024charge}. Addressing the competition between fractionalization and crystallization in an unbiased manner remains an outstanding challenge, because an interacting 2DEG in a magnetic field features an infinite ladder of Landau levels and strong electron correlation. Quantum Monte Carlo faces severe complex phase problem; and density matrix renormalization group suffers from discretization errors due to Landau level truncation. As a result, much of our understanding relies on trial wave functions that are tailor made for FQH liquids and Wigner crystals separately, making it difficult to determine the phase boundary in an unbiased way.

In recent years, neural-network quantum states \cite{carleo2017solving,pfau2020ab,luo2019backflow}  have emerged as a powerful new class of variational wave functions and have attained accurate results in continuum Fermi systems \cite{cassella2023discovering, smith2024unified, pescia2024message, luo2023pairing}. Here the networks are trained directly by minimizing the variational energy of the microscopic Hamiltonian. Remarkably, recent studies \cite{teng2025solving, geier2025self,li2025attention,nazaryan2025artificial, abouelkomsan2025topological} have found that ground states of vastly different quantum phases---ranging from Fermi liquids and Wigner crystals to fractional Hall states and superconductors---can all be captured within a single neural architecture; fundamentally distinct quantum states simply correspond to different values of network parameters. 

This unprecedented expressive power has further motivated the development of {\it universal} Fermi networks that are provably capable of representing any fermionic wave function at sufficient network size \cite{fu2025minimal}. Recently, a universal architecture---``Fermi Sets''---has been introduced \cite{fu2026fermi}, which is mathematically proven to be universal approximators of continuous fermionic wave functions while retaining physical interpretability. Universal Fermi networks of sufficiently large size can in principle solve many-electron Schrodinger equations to arbitrary accuracy. This opens vast opportunities for {\it first-principles AI} in quantum chemistry, condensed matter physics, material science and quantum computing.

In this work, we develop a self-attention Fermi network (MagNet) to solve the strongly correlated problem of two-dimensional interacting electrons in a magnetic field, where fractionalization competes with crystallization. In contrast to previous neural-network variational studies \cite{qian2025describing,teng2025solving}, we work on the torus geometry, which is free of boundary effects and naturally accommodates both topological fluids and crystalline order in an unbiased way. A key innovation is our construction of a real-space neural wave function that exactly respects the nontrivial boundary conditions imposed by magnetic translations while remaining extremely expressive. In particular, our network allows for general, intricate phase structures far beyond those of standard quantum Hall model wave functions.

\begin{figure*}[t!]
    \centering
    \includegraphics[width=0.9\linewidth]{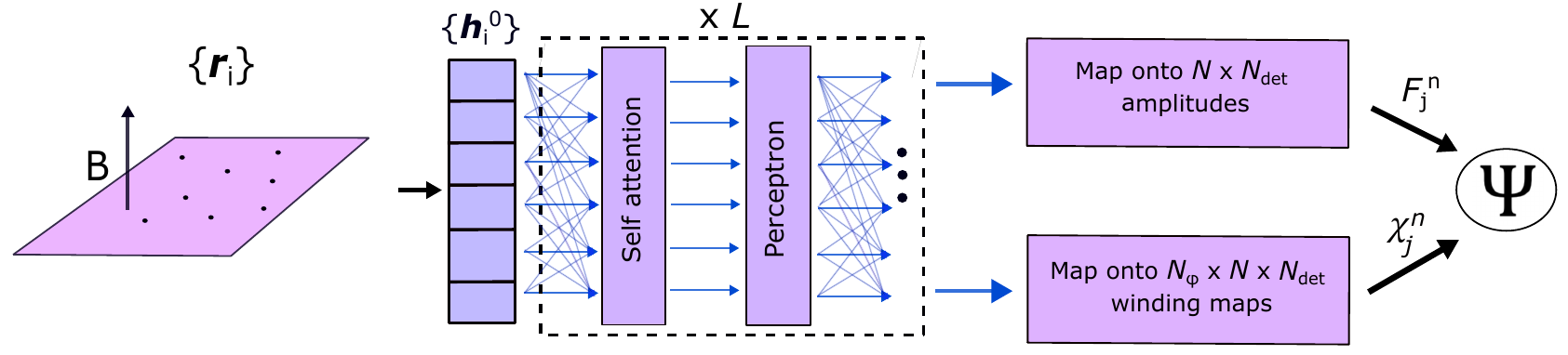}
    \caption{Schematic illustration of the MagNet architecture: Electron coordinates (in the presence of magnetic field) are mapped to higher dimensional feature space which is passed to $L$ layers of successive self-attention and multi-layer perceptron blocks. The final output undergo two different mappings to form the generalized orbitals \eqref{eq:phiproduct} which build the full variational ansatz. For simplicity, we show a plane pierced by a magnetic field but our ansatz is designed for the torus geometry. }
    \label{fig:schematic}
\end{figure*}

Using this unifying architecture, we obtain accurate ground-state energies and wave functions across the entire range of Landau-level (LL) mixing, from weak to strong. The same neural network discovers both FQH liquids and electron crystals directly from the microscopic Hamiltonian, without being supplied with any physics knowledge, such as Landau levels, Laughlin states, flux attachment, or crystalline order. All information about fractionalization and crystallization is extracted a posteriori from the wavefunction learned by the network through energy minimization alone. In this sense, our work realizes a genuinely first-principles AI solver for this paradigmatic strongly correlated problem. 

Beyond verifying known phases in the limit of weak and strong LL mixing, our work provides, for the first time, a unified  solution across the FQH-to-crystal topological quantum phase transition, within a single family of variational wavefunctions. This enables us to follow the evolution of correlation functions and structure factors and locate the onset of crystalline order.

More broadly, our study highlights the power of first-principles AI as a general-purpose tool for quantum matter, capable of exploring phase diagrams, discovering unexpected correlated phases, and offering new microscopic insights into the organizing principles of strongly interacting electrons.

\par

\textit{Setup---}  Our starting point is the standard problem of two dimensional electrons subject to an external magnetic field and interacting with Coulumb potential. The many-body Hamiltonian of $N$ particles reads,

\begin{equation}
\label{eq:Ham}
    H = \sum_{i} \dfrac{(-i \hbar \boldsymbol{\nabla}_i + e \A(\r_i))^2}{2 m} + \frac{1}{2} \sum_{i \neq j} \frac{e^2}{4 \pi \epsilon |\r_i - \r_j|}
\end{equation}
where $\A(\r)$ is the vector potential of the magnetic field $\boldsymbol{B} = \nabla \cross \A(\r) $. This problem is governed by two energy scales, the kinetic energy scale $ K = \hbar \omega_c$ which sets the gap between the (infinite) ladder of flat Landau levels with $\omega_c = eB/m$ the cyclotron frequency and interaction scale $U = e^2/4\pi \epsilon \ell_B$ which sets the strength of the Coulomb repulsion in terms of the magnetic length $\ell_B = \sqrt{\hbar/eB}$. The many-body ground state at filling factor $\nu$ is therefore controlled by the ratio of the two, $\kappa = U/K\propto 1/\sqrt{B}$ which describes the amount of Landau level mixing. $\kappa$ is related to the dimensionless interaction strength parameter $r_s = 1/\sqrt{\pi a_B^2 n}$ through $\kappa = r_s \sqrt{\nu/2}$ where $a_B = \hbar^2/e^2m$ is the Bohr radius and $n$ is the density. Experimentally, for a given magnetic field, $\kappa$ is material-dependent and can become very large in materials with heavy effective mass, such as hole doped GaAs, ZnO or transition metal dichalcogenides (TMDs) \cite{sodemann2013landau,wang2023next,maryenko2018composite,tsukazaki2010observation,pack2024charge,shi2020odd}.

Focusing on $\nu < 1$, the physics in the limit $\kappa \rightarrow 0$ is confined to the lowest Landau level where fractional quantum Hall liquids arise at various fillings, while increasing $\kappa$ drives substantial Landau-level mixing and can  quantitatively change the ground state. In particular, when $\kappa \rightarrow \infty$, interactions dominate over kinetic energy and the ground state becomes a classical Wigner crystal. At intermediate values of $\kappa$, the system lies in a genuinely nonperturbative regime, 
and the FQH liquid competes with various possible charge-ordered states such as Wigner crystals, composite fermion crystals \cite{archer2011static,yi1998laughlin,narevich2001hamiltonian,chang2005microscopic,archer2013competing,zhao2018crystallization}, or Hall crystals \cite{tevsanovic1989hall}. Resolving this competition requires a method that can treat strong Landau-level mixing, while remaining unbiased with respect to competing states.

To proceed, we work on the torus geometry which is defined by two basis vectors $\L_1$ and $\L_2$ and is pierced by a magnetic field whose total flux is quantized to an integer number $N_\phi$ of flux quanta, fixed by $|\L_1 \cross \L_2| = 2\pi N_\phi \ell_B^2$. The torus provides a particularly favorable geometry for studying competing quantum phases.  Unlike the disk, it has no physical edge and therefore avoids boundary effects and edge reconstruction \cite{wan2003edge}. Compared to the sphere, it also avoids curvature and the associated topological shift \cite{wen1992shift} that can complicate finite-size comparisons across different phases. Additionally, the torus naturally accommodates translational symmetry, making it well suited not only for uniform quantum Hall liquids but also for charge-ordered phases such as Wigner crystals, whose periodicity can be embedded commensurately in the torus supercell.

\textit{MagNet} --- 
To comply with the periodicity of the torus, any wavefunction has to transform under translation of a single particle as \cite{supplementary},
\begin{equation}
\label{eq:magBC}
    \Psi(\r_1, \cdots, \r_i + \L, \cdots, \r_N) = e^{i\varphi} e^{i \xi_{\L}(\r_i)}  \Psi(\r_1, \cdots, \r_N)
\end{equation}
where $\xi_{\L}(\r_i)$ is the magnetic phase determined by $\nabla \xi_{\L}(\r_i) = \frac{e}{\hbar} [\A (\r_i + \L) - \A(\r_i)]$ and $\L = m \L_1 + n \L_2$ is a generic torus vector with integers $m$ and $n$. $\varphi$ is an additional twist phase corresponding to threading flux through the handles of the torus,  implemented by a constant vector potential.

Importantly, the condition \eqref{eq:magBC} implies that the net winding $\mathcal{W}$ of the wavefunction as a function of one particle coordinate $\r_i$ (while fixing the remaining $N-1$ particles $\{\r_{\neq i}\}$) has to be $ \mathcal{W} = N_\phi$ \cite{haldanemanybody,haldanerezayi1985}. This net winding constraint requires the wavefunction to contain phase singularities on the torus---i.e., vortices where $\Psi$ vanishes and its phase winds. In general, the wavefunction may host both vortices and antivortices, but their contributions must sum to the same net winding: the total winding (vortices counted with positive charge minus antivortices counted with negative charge) is fixed to $N_\phi$.

 Our variational wavefunction for the many-body ground state is given by,
\begin{equation}
\label{eq:ansatz}
    \Psi_{\{\theta\}}(\{\r_i\})  = e^{\mathcal{J}(\{\r_i\})} \sum_{n}^{N_{\rm det}} {\rm det}[\phi_j^n(\r_i; \{\r_{\neq i }\})]
\end{equation} where $\phi_j^n(\r_i; \{\r_{\neq i }\})$ is a generalized many-body orbital for the $i$-th particle that depends on the coordinates of the remaining particles $\r_{\neq i }$. $\{\theta\}$ denotes the variational parameters of the ansatz. $\mathcal{J}(\{\r_i\})$ is a periodic symmetric Jastrow factor , $\mathcal{J}(\r_1, \cdots, \r_i + \L, \cdots, \r_N) = \mathcal{J}(\r_1,\cdots,\r_N)$, to enforce electron cusp conditions for long-range Coulomb interactions. However, the existence of a Jastrow factor here is not generally needed and we obtained similar results without it. 

To respect fermionic antisymmetry, $\phi_j^n$ has to be permutation \textit{invariant} in the coordinates $\{\r_{\neq i }\}$ \cite{pfau2020ab,geier2025self}. Moreover, to enforce magnetic boundary condition $\eqref{eq:magBC}$, $\phi_j^n$ transforms under translations of $\r_i$ as,
\begin{equation}
\label{eq:quasiperiodic}
     \phi_j^n(\r_i + \L, \{\r_{\neq i }\}) = e^{i \varphi} e^{i\xi_{\L}(\r_i)}  \phi_j^n(\r_i; \{\r_{\neq i }\})
\end{equation}
and is periodic in the rest $\{\r_{\neq i }\}$
\begin{equation}
\label{eq:periodic}
     \phi_j^n(\r_i, \{\dots, \r_k + \L, \dots,\r_l, \dots\}) = \phi_j^n(\r_i; \{\r_{\neq i }\})
\end{equation}
for $k,l, \dots, \neq i$. A sum of determinants over $\phi_j^n$ as in the ansatz \eqref{eq:ansatz} automatically satisfies the condition \eqref{eq:magBC} and therefore represents a physically valid wavefunction on the torus. 

Furthermore, $\phi_j^n$ is factorized into a product, \begin{equation}
\label{eq:phiproduct}
    \phi_j^n(\r_i; \{\r_{\neq i }\}) = \chi_j^n(\r_i; \{\r_{\neq i }\}) F_j^n(\r_i; \{\r_{\neq i }\})
\end{equation}
where $F_j^n$ is a periodic function in all coordinates, $F_j^n(\r_1, \dots, \r_i + \L, \r_N) = F_j^n(\r_1, \cdots, \r_N)$ while $\chi_j^n$ transforms similar to $\phi_j^n$ (Equations \eqref{eq:quasiperiodic} and \eqref{eq:periodic}). 

To impose a net winding of $N_\phi$, the core innovation of our ansatz is to parametrize $\chi_j^n$ in terms of a product over $N_\phi$ terms,
\begin{equation}
\label{eq:zeroparam}
    \chi_j^n(\r_i; \{\r_{\neq i }\}) = \prod_{\alpha= 1}^{N_\phi} f(z_i - \eta^{(n ,\alpha)}_{j}(\{\r\})) 
\end{equation}
where $f(z_i)$ is a gauge-dependent quasi-periodic function in the coordinate $z_i$ with net winding $\mathcal{W} = +1$ in the fundamental domain of the torus spanned by $\L_1$ and $\L_2$ and $z_i = x_i + i y_i$ is the complex coordinate of $\r_i$. For our purposes, the function $f(z_i - \eta)$ vanishes when $z_i = \eta$. Importantly, $\eta^{(n,\alpha)}_{j}(\{\r\})$ here is a \textit{learnable} periodic and symmetric many-body function  in \textit{all} coordinates.  The zeros of $\chi_j^n$ and their windings are determined through the solutions of $z_i = \eta^{(n,\alpha)}_{j}(\{\r\})$. For this reason, we refer to $\eta_j^{(n,\alpha}(\{\r\})$ as a winding map.

We note that if the winding maps $\{\eta\}$ in \eqref{eq:zeroparam} are taken to be fixed parameters (no dependence on $\{\r\}$), then $\chi_j^{n}$ reduces to the single particle lowest Landau level (LLL) wavefunction which is specified by  $N_\phi$ vortices and is holomorphic (up to a Gaussian factor) \cite{haldanemanybody,fremling2013coherent}. In contrast, allowing $\{\eta\}$ to depend on all particle coordinates---including $\r_i$ itself---generically introduces $\bar z$-dependence, so that $\chi_j^{n}$ becomes non-holomorphic. Together with the additional periodic factor $F_j^{n}$, this non-holomorphic structure enables the full many-body ansatz to represent states residing outside the LLL. Crucially, by promoting the zeros of the generalized orbitals to be many-body trainable functions, the variational ansatz can capture non-trivial phase effects in a correlated manner.

\begin{figure}[t!]
    \centering
    \includegraphics[width=\linewidth]{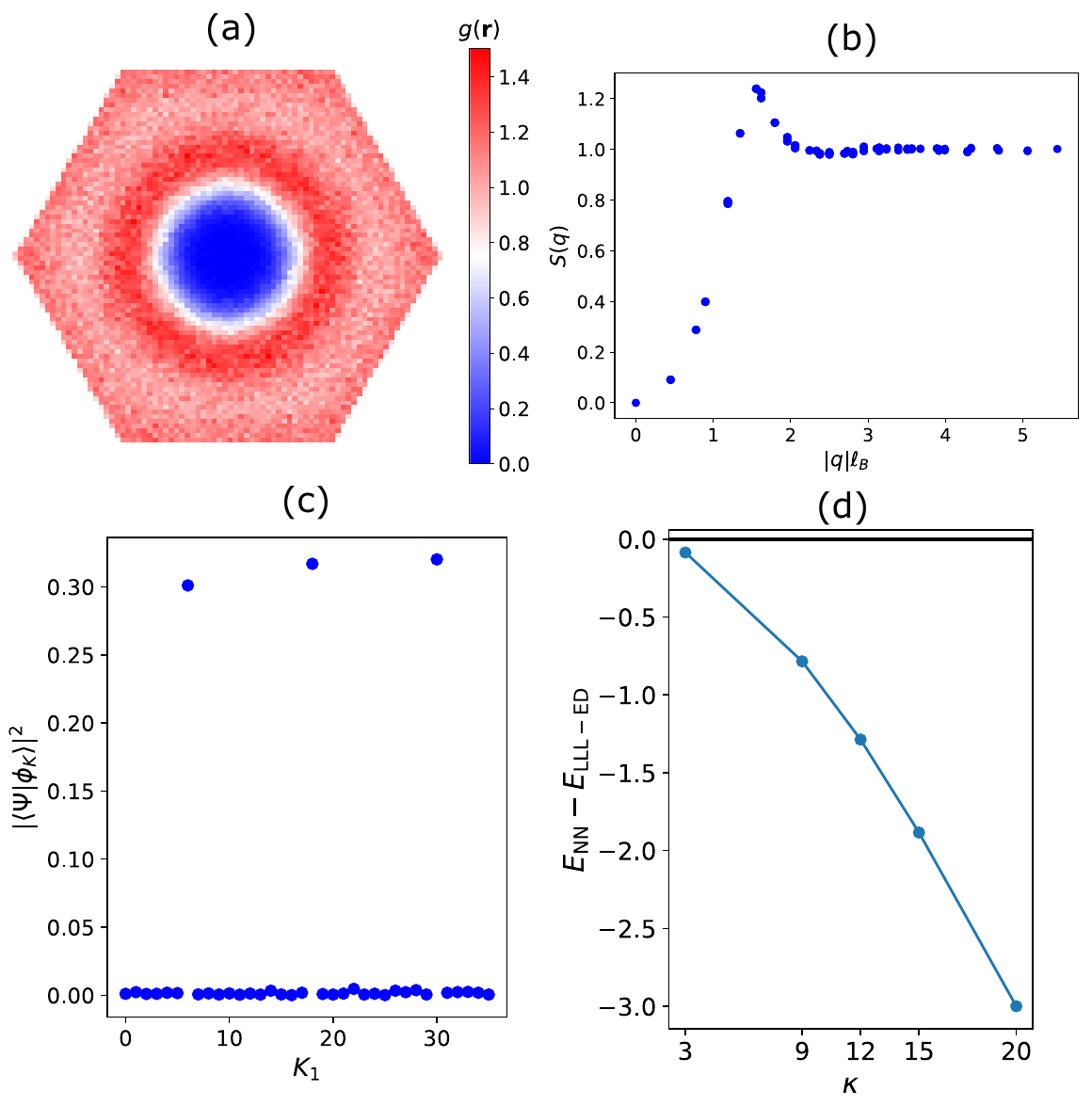}
    \caption{(a) Pair correlation function $g(\r)$ at $\kappa = 3.0$. (b) The structure factor $S(\q)$ along a line cut at $\kappa = 3.0$. (c) The overlap $|\langle \Psi|\Phi_{K_1} \rangle|^2$ of the optimized wavefunction $\Psi$ at $\kappa = 3.0$ with its projection $\Phi_{K_1}$ defined as the eigenstates of CM magnetic translation operator along $L_1$. (d) Comparison of the variational energy of the NN ansatz $E_{\rm NN}$ (in units of $\hbar \omega_c$) against the energies obtained from exact diagonalization $E_{\rm LLL-ED}$ projected onto the lowest Landau level for various values of $\kappa$.
    All calculations are performed at filling $\nu=1/3$ ($N = 12$ and $N_{\phi} = 36$) on a hexagonal torus with equal aspect ratio. }
    \label{fig:fqhdemonstration}
\end{figure}

\begin{figure*}[t!]
    \centering
    \includegraphics[width=\linewidth]{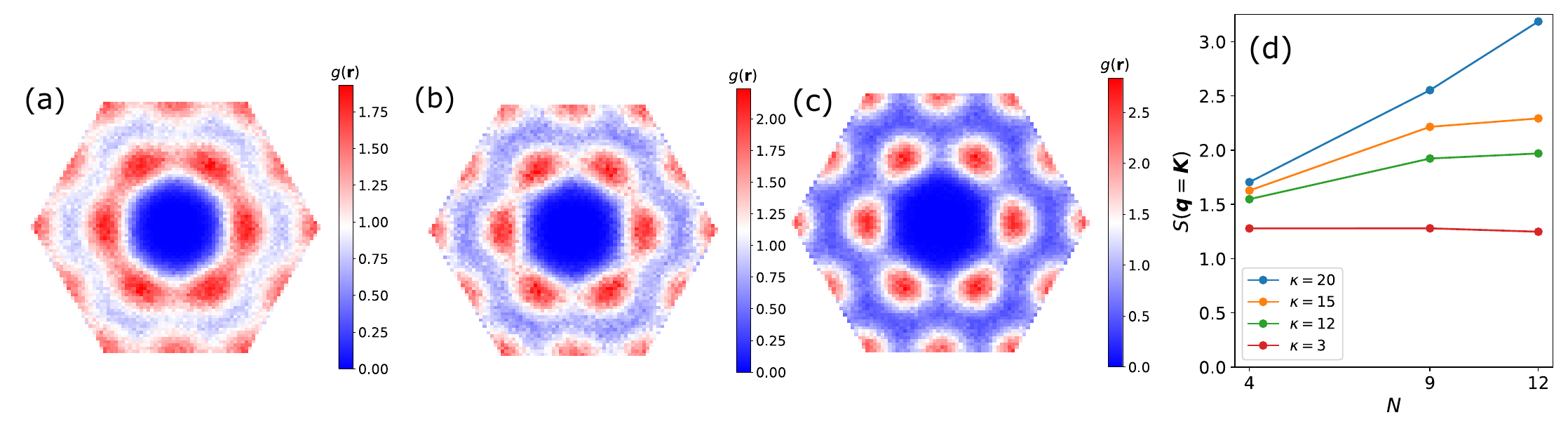}
    \caption{Pair correlation function $g(\r)$ at $\nu = 1/3$ for (a) $\kappa = 12$ , (b) $\kappa = 15$ and (c) $\kappa = 20$. (d) Structure factor $S(\q)$ evaluated at the ordering wavevector $\q=\K$ corresponding to a crystal with one electron per unit cell, plotted as a function of particle number $N$ for various values of $\kappa$. The reported $S( \q = \K)$ is rotationally averaged over the six $C_6$-related ordering wavevectors. All calculations are performed with $N = 12$ and $N_\phi = 36$ on a hexagonal torus with equal aspect ratio}
    \label{fig:paircorrelations_withkappa}
\end{figure*}

To proceed, we work in the symmetric gauge,
$A(\r) = B_0 (\hat{z} \cross \r)/2$,
where $B_0$ denotes the uniform component of the magnetic field. In this gauge, the magnetic translation phase takes the form
\begin{equation}
    e^{i\xi_{\L}(\r)} = (-1)^{N_\phi (m n + m + n)} e^{i (\r \cross \L) / 2\ell_B^2}
\end{equation}
for $ \L = m \L_1 + n \L_2$. We choose $f(z)$ in Eq.~\eqref{eq:zeroparam} to be the modified Weierstrass sigma function \cite{haldane2018modular,haldane2018origin,wang2019lattice}. In this representation, the sum of the winding maps,
\(\overline{\eta}^{(n)}_j=\big(\sum_{\alpha}\eta^{(n,\alpha)}_j\big)\bmod(L_x+iL_y)\),
fixes the twist phase $\varphi$ in Eq.~\eqref{eq:quasiperiodic}. Throughout, we set $\varphi=0$ by imposing the constraint $\overline{\eta}^{(n)}_j=0$ \cite{supplementary}.

Next, we represent the generalized orbitals $\phi_j^n$ with a deep neural network (Fig. \ref{fig:schematic}) based on the self-attention mechanism \cite{von2022self,geier2025self}. As initial input, the NN takes the \textit{periodized} coordinates of the particles on the torus [$\sin(\G_a \cdot \r_i)$ and $\cos(\G_a \cdot \r_i)$] where $\G_{a = 1,2}$ are the reciprocal lattice vectors, i.e,  $\G_a \cdot \L_b = 2\pi \delta_{ab} $. These inputs are then passed through $L$ layers of successive self-attention and perceptron blocks. The final output of the NN is mapped to construct the amplitudes $F_j^n$ and the winding maps $\eta^{(n,\alpha)}_j$ of $\chi_j^n$ to form the generalized orbitals $\phi_j^n$. The parameters of the ansatz $\{\theta\}$ are optimized through energy minimization to yield the final ansatz \eqref{eq:ansatz}.  Our construction is distinct from very recent work \cite{fadon2025extracting} in several key aspects. Rather than a linear combination of single particle orbitals, the quasi-periodic function $\chi_j^n$ \eqref{eq:zeroparam}  is parametrized by its zeros as determined by winding maps $\{\eta\}$ that depend on \textit{all} particle coordinates, leading to enhanced expressivity. Moreover, as we shall demonstrate below, our parametrization successfully captures fractional quantum Hall liquids without additional neural-network Jastrow factor. We refer the reader to the supplemental material \cite{supplementary} for comprehensive details about the architecture and the optimization.

\textit{Results ---} To test the validity of our ansatz, we apply it to the Hamiltonian \eqref{eq:Ham} at filling factor $\nu = 1/3$. At $\kappa=3.0$, the optimized neural-network wavefunction exhibits hallmarks of a fractional quantum Hall liquid.
Fig. \ref{fig:fqhdemonstration}(a) shows the pair correlation function $g(\r) = \frac{A}{N^2} \langle \sum_{i \neq j} \delta^2(\r - \r_i + \r_j) \rangle $ and Fig. \ref{fig:fqhdemonstration}(b) shows the structure factor $S(\q) = \frac{1}{N} \langle \rho_{\q} \rho_{-\q} \rangle$ with the density operator $\rho_{\q} = \sum_{i} e^{i \q \cdot \r_i}$. In both cases, the density correlations in the ground state are liquid-like indicating the absence of crystalline order in the ground state. 

Next, we probe the existence of topological order by demonstrating the expected topological ground state degeneracy \cite{abouelkomsan2025topological}. To this end, we decompose the optimized many-body wavefunction $\Psi(\{\r\})$ onto center of mass (CM) momentum sectors $\Phi_{K}(\{\r\})$, defined as eigenstates of the CM magnetic translation operator $T_{\rm CM}(n \L_1/N_\phi)$ with $n = 0, \cdots, N_\phi - 1$. As shown in Fig. \ref{fig:fqhdemonstration}(c), the resulting state carries nonzero weight in only three CM momentum sectors, matching precisely the sectors of the $\nu = 1/3$ Laughlin state for the same finite-size geometry. Taken together, our findings establishes the ability of the NN ansatz to discover a fractional quantum Hall liquid on the torus geometry. 

To further corroborate the effectiveness of our ansatz in capturing strong Landau-level mixing, we compare its energy against lowest-Landau-level (LLL) projected exact diagonalization (ED) in Fig.~\ref{fig:fqhdemonstration}(d). We find that the NN achieves consistently lower variational energies than the corresponding  ED energies. This highlights a key advantage of our approach: for this system size ($N_\phi = 36$), ED is restricted to the LLL and can only capture a FQH liquid regardless of $\kappa$, whereas our real-space NN ansatz naturally incorporates Landau-level mixing without truncation and allows for competing phases. 

To address whether strong Landau level mixing can destabilize the fractional quantum Hall liquid, we use our unified NN ansatz to study the evolution of the many-body ground state as a function of $\kappa$. As shown in Fig.~\ref{fig:paircorrelations_withkappa}(a)-(c), the pair-correlation function $g(\r)$ evolves from being featureless and liquid-like to exhibiting well-developed spatial modulations as $\kappa$ increases, signaling a transition toward a crystalline phase. 

To assess whether these correlations correspond to true long-range crystalline order in the thermodynamic limit, we perform a finite-size scaling of the structure factor $S(\q)$. First, we find $S(\q)$ to have maximum value at $\q = \K$ for various values of $\kappa$ with $\K$ being the ordering wavevector of a crystal with one electron per unit cell. However, as shown in Fig.~\ref{fig:paircorrelations_withkappa}(d), the magnitude of this peak $S(\K)$ exhibits only a small enhancement with increasing particle number $N$ for $\kappa=12$ and $\kappa=15$, suggesting at most incipient charge order at these values. In contrast, for $\kappa=20$ we find that $S(\K)$ grows approximately linearly with $N$, consistent with the development of true long-range crystalline order. On the other hand, for $\kappa = 3$, $S(\K)$ does not grow with $N$ and is nearly constant, consistent with a FQH liquid.

\begin{figure}
    \centering
    \includegraphics[width=0.9\linewidth]{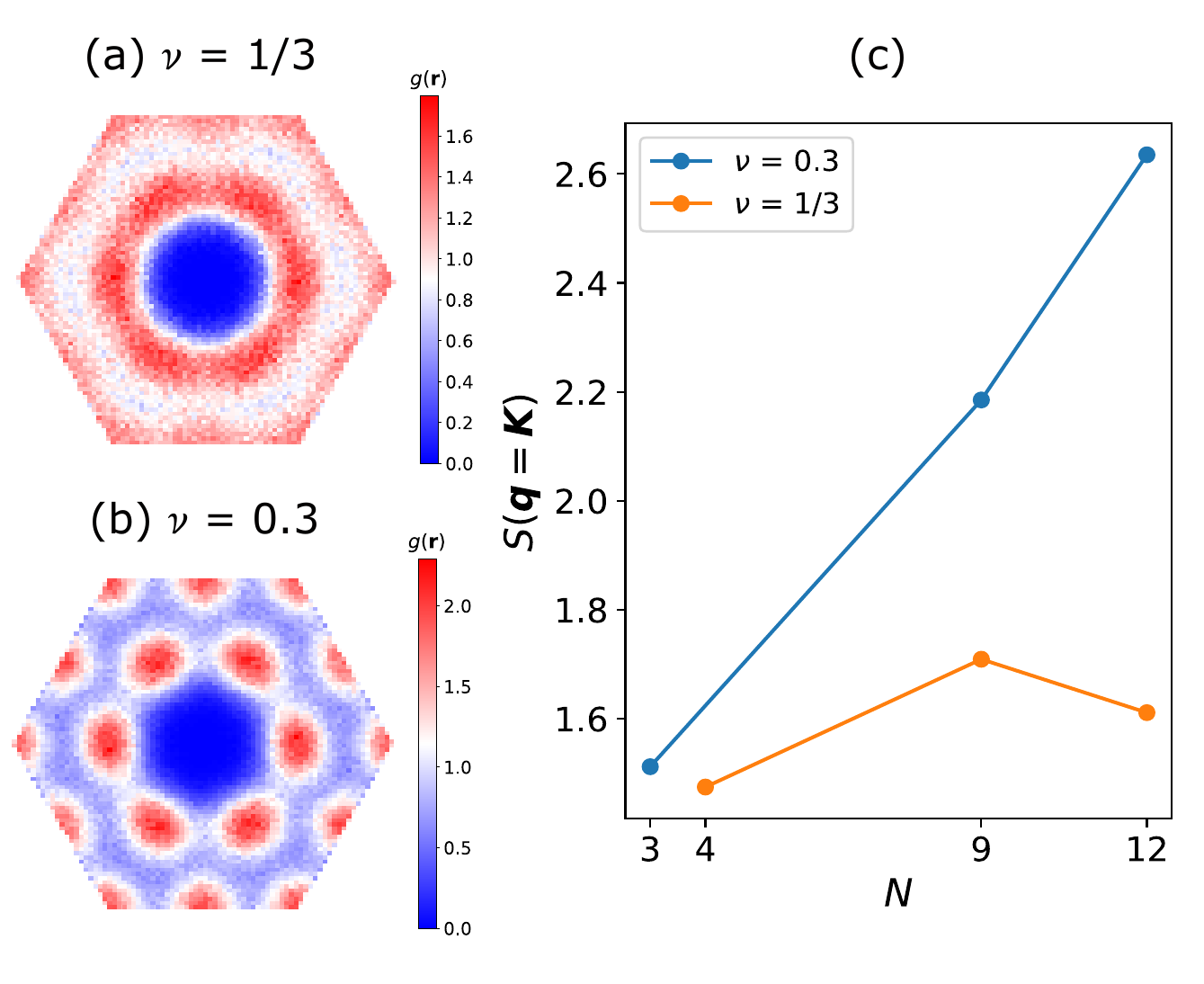}
    \caption{{(a)-(b) Pair correlation function $g(\r)$ at $\kappa = 9$ for (a) $\nu = 1/3$ and (b) $\nu = 0.3$. (c)  Structure factor $S(\q)$ evaluated at the ordering wavevector $\q=\K$ corresponding to a crystal with one electron per unit cell as a function of number of particles $N$  at $\kappa = 9$. Calculations in (a) and (b) are performed with $N = 12$ on a hexagonal torus enclosing $N_\phi = 36$ and $N_\phi = 40$.}}
    \label{fig:maxsqvsN}
\end{figure}

The robustness of the Laughlin liquid at $\nu=1/3$ against crystallization for strong Landau level mixing is rather special, as can be seen by contrasting it with nearby fillings. As shown in Fig. \ref{fig:maxsqvsN}, at the same value of Landau-level mixing for which the $\nu=1/3$ ground state remains a homogeneous liquid [Fig. \ref{fig:maxsqvsN}(a)], a slight doping away from this commensurate filling, to $\nu=0.3$, already favors a crystalline ground state [Fig. \ref{fig:maxsqvsN}(b)]. This is further supported by the finite-size scaling of the structure factor shown in Fig. \ref{fig:maxsqvsN}(c). The existence of crystallization in the vicinity of the FQH liquid is consistent with experimental results \cite{santos1992observation,santos1992effect,maryenko2018composite,pan2005transition}.
 
\textit{Discussion ---} In this paper, we have introduced MagNet, a self-attention neural network variational wavefunction to study quantum systems subject to magnetic fields on the torus geometry. We have shown that MagNet captures both fractional quantum Hall liquids and crystals in a unified manner.  While we have presented evidence of pronounced crystalline correlations with increasing Landau level mixing, the full nature of the crystal phase is yet to be established. Apart from topologically trivial Wigner crystal or the composite fermion crystal, an intriguing possibility is the realization of (integer or fractional) Hall crystal \cite{tevsanovic1989hall,halperin1986compatibility,murthy2000hall,kivelson1986cooperative}, a state that spontaneously break magnetic translational symmetry but has quantized Hall conductance.

Earlier numerical work \cite{price1993fractional} suggested that the $\nu=1/3$ FQH liquid undergoes a transition to a crystalline phase at  Landau-level mixing, around $\kappa \approx 9$ (corresponding to $r_s \approx 22$). In contrast, more recent studies based on fixed-phase diffusion Monte Carlo (DMC) \cite{zhao2018crystallization}---where the complex phase of the trial state is constrained to be that of a composite-fermion wavefunction---did not find crystallization of the $\nu=1/3$ FQH liquid up to the largest mixing strength considered, $\kappa=18$ ($r_s \approx 44$). In comparison with these results, our unbiased variational Monte Carlo results---obtained within a single unified ansatz and based on the system sizes accessible here---indicate that long-range crystalline order emerges between $\kappa=15$ ($r_s \approx 37$) and $\kappa=20$ ($r_s \approx 49$) (cf.~Fig.~\ref{fig:maxsqvsN}). 

The development of a neural-network ansatz that can treat arbitrary Landau-level mixing opens a powerful route to studying a wide range of quantum Hall problems in an unbiased, first-principles manner. While we have focused on $\nu=1/3$, a natural next step is to apply this framework across a broader set of filling factors to investigate the competition between liquid phases---including incompressible FQH states and compressible composite-fermion Fermi liquids---and crystalline order.

Finally, our neural-network ansatz is naturally suited to problems with spatially non-uniform magnetic fields or external modulations \cite{paul2023giant}, such as moir\'e systems---for example, twisted transition-metal dichalcogenides---where the moir\'e potential generates emergent spatially modulated magnetic fields \cite{wu2019topological,morales2024magic,shi2024adiabatic}.

\textit{Acknowledgements---} We are grateful to  Aidan Reddy, Timothy Zaklama and Filippo Gaggioli for useful discussions and related collaborations. We thank Timothy Zaklama for coining the NN architecture name and acknowledge insightful discussions with Max Geier, Daniele Guerci and Pierre-Antoine Graham. This work was supported by a Simons Investigator Award from the Simons Foundation. A. A. was
supported by the Knut and Alice Wallenberg Foundation
(KAW 2022.0348). 
This work made use of resources provided by subMIT at MIT Physics. The authors acknowledge the MIT Office of Research Computing and Data for providing high performance computing resources that have contributed to the research results reported within this paper.
\bibliography{references}

@article{pfau2020ab,
  title={Ab initio solution of the many-electron Schr{\"o}dinger equation with deep neural networks},
  author={Pfau, David and Spencer, James S and Matthews, Alexander GDG and Foulkes, W Matthew C},
  journal={Physical review research},
  volume={2},
  number={3},
  pages={033429},
  year={2020},
  publisher={APS}
}

@article{von2022self,
  title={A self-attention ansatz for ab-initio quantum chemistry},
  author={von Glehn, Ingrid and Spencer, James S and Pfau, David},
  journal={arXiv preprint arXiv:2211.13672},
  year={2022}
}

@article{geier2025self,
  title={Self-attention neural network for solving correlated electron problems in solids},
  author={Geier, Max and Nazaryan, Khachatur and Zaklama, Timothy and Fu, Liang},
  journal={Physical Review B},
  volume={112},
  number={4},
  pages={045119},
  year={2025},
  publisher={APS}
}

@article{li2025attention,
  title={Attention is all you need to solve chiral superconductivity},
  author={Li, Chun-Tse and Ong, Tzen and Geier, Max and Lin, Hsin and Fu, Liang},
  journal={arXiv preprint arXiv:2509.03683},
  year={2025}
}

@misc{supplementary,
	Note = {see Supplementary Material for additional details.},
}

@article{haldanerezayi1985,
  title = {Periodic Laughlin-Jastrow wave functions for the fractional quantized Hall effect},
  author = {Haldane, F. D. M. and Rezayi, E. H.},
  journal = {Phys. Rev. B},
  volume = {31},
  issue = {4},
  pages = {2529--2531},
  numpages = {0},
  year = {1985},
  month = {Feb},
  publisher = {American Physical Society},
  doi = {10.1103/PhysRevB.31.2529},
  url = {https://link.aps.org/doi/10.1103/PhysRevB.31.2529}
}

@article{haldane2018modular,
  title={A modular-invariant modified Weierstrass sigma-function as a building block for lowest-Landau-level wavefunctions on the torus},
  author={Haldane, FDM},
  journal={Journal of Mathematical Physics},
  volume={59},
  number={7},
  year={2018},
  publisher={AIP Publishing}
}

@article{haldane2018origin,
  title={The origin of holomorphic states in Landau levels from non-commutative geometry and a new formula for their overlaps on the torus},
  author={Haldane, FDM},
  journal={Journal of Mathematical Physics},
  volume={59},
  number={8},
  year={2018},
  publisher={AIP Publishing}
}

@article{wang2019lattice,
  title={Lattice Monte Carlo for quantum Hall states on a torus},
  author={Wang, Jie and Geraedts, Scott D and Rezayi, EH and Haldane, FDM},
  journal={Physical Review B},
  volume={99},
  number={12},
  pages={125123},
  year={2019},
  publisher={APS}
}

@article{abouelkomsan2025topological,
  title={Topological Order in Deep State},
  author={Abouelkomsan, Ahmed and Geier, Max and Fu, Liang},
  journal={arXiv preprint arXiv:2512.01863},
  year={2025}
}

@article{shi2020odd,
  title={Odd-and even-denominator fractional quantum Hall states in monolayer WSe2},
  author={Shi, Qianhui and Shih, En-Min and Gustafsson, Martin V and Rhodes, Daniel A and Kim, Bumho and Watanabe, Kenji and Taniguchi, Takashi and Papi{\'c}, Zlatko and Hone, James and Dean, Cory R},
  journal={Nature Nanotechnology},
  volume={15},
  number={7},
  pages={569--573},
  year={2020},
  publisher={Nature Publishing Group UK London}
}

@article{haldanemanybody,
  title = {Many-Particle Translational Symmetries of Two-Dimensional Electrons at Rational Landau-Level Filling},
  author = {Haldane, F. D. M.},
  journal = {Phys. Rev. Lett.},
  volume = {55},
  issue = {20},
  pages = {2095--2098},
  numpages = {0},
  year = {1985},
  month = {Nov},
  publisher = {American Physical Society},
  doi = {10.1103/PhysRevLett.55.2095},
  url = {https://link.aps.org/doi/10.1103/PhysRevLett.55.2095}
}

@article{teng2025solving,
  title={Solving the fractional quantum Hall problem with self-attention neural network},
  author={Teng, Yi and Dai, David D and Fu, Liang},
  journal={Physical Review B},
  volume={111},
  number={20},
  pages={205117},
  year={2025},
  publisher={APS}
}

@article{nazaryan2025artificial,
  title={Artificial Intelligence for Quantum Matter: Finding a Needle in a Haystack},
  author={Nazaryan, Khachatur and Gaggioli, Filippo and Teng, Yi and Fu, Liang},
  journal={arXiv preprint arXiv:2507.13322},
  year={2025}
}

@article{fu2026fermi,
  title={Fermi Sets: Universal and interpretable neural architectures for fermions},
  author={Fu, Liang},
  journal={arXiv preprint arXiv:2601.02508},
  year={2026}
}

@article{fu2025minimal,
  title={A minimal and universal representation of fermionic wavefunctions (fermions= bosons+ one)},
  author={Fu, Liang},
  journal={arXiv preprint arXiv:2510.11431},
  year={2025}
}

@article{sodemann2013landau,
  title={Landau level mixing and the fractional quantum Hall effect},
  author={Sodemann, Inti and MacDonald, AH},
  journal={Physical Review B—Condensed Matter and Materials Physics},
  volume={87},
  number={24},
  pages={245425},
  year={2013},
  publisher={APS}
}

@article{wang2023next,
  title={Next-generation even-denominator fractional quantum Hall states of interacting composite fermions},
  author={Wang, Chengyu and Gupta, Adbhut and Madathil, Pranav T and Singh, Siddharth K and Chung, Yoon Jang and Pfeiffer, Loren N and Baldwin, Kirk W and Shayegan, Mansour},
  journal={Proceedings of the National Academy of Sciences},
  volume={120},
  number={52},
  pages={e2314212120},
  year={2023},
  publisher={National Academy of Sciences}
}

@article{zhao2018crystallization,
  title={Crystallization in the fractional quantum Hall regime induced by Landau-level mixing},
  author={Zhao, Jianyun and Zhang, Yuhe and Jain, JK},
  journal={Physical review letters},
  volume={121},
  number={11},
  pages={116802},
  year={2018},
  publisher={APS}
}

@article{luo2019backflow,
  title={Backflow transformations via neural networks for quantum many-body wave functions},
  author={Luo, Di and Clark, Bryan K},
  journal={Physical review letters},
  volume={122},
  number={22},
  pages={226401},
  year={2019},
  publisher={APS}
}

@article{ortiz1993new,
  title={New stochastic method for systems with broken time-reversal symmetry: 2D fermions in a magnetic field},
  author={Ortiz, G and Ceperley, DM and Martin, RM},
  journal={Physical review letters},
  volume={71},
  number={17},
  pages={2777},
  year={1993},
  publisher={APS}
}

@article{maryenko2018composite,
  title={Composite fermion liquid to Wigner solid transition in the lowest Landau level of zinc oxide},
  author={Maryenko, D and McCollam, A and Falson, J and Kozuka, Y and Bruin, J and Zeitler, U and Kawasaki, M},
  journal={Nature communications},
  volume={9},
  number={1},
  pages={4356},
  year={2018},
  publisher={Nature Publishing Group UK London}
}

@article{archer2013competing,
  title={Competing crystal phases in the lowest Landau level},
  author={Archer, Alexander C and Park, Kwon and Jain, Jainendra K},
  journal={Physical review letters},
  volume={111},
  number={14},
  pages={146804},
  year={2013},
  publisher={APS}
}

@article{narevich2001hamiltonian,
  title={Hamiltonian theory of the composite-fermion Wigner crystal},
  author={Narevich, R and Murthy, Ganpathy and Fertig, HA},
  journal={Physical Review B},
  volume={64},
  number={24},
  pages={245326},
  year={2001},
  publisher={APS}
}

@article{yi1998laughlin,
  title={Laughlin-Jastrow-correlated Wigner crystal in a strong magnetic field},
  author={Yi, Hangmo and Fertig, HA},
  journal={Physical Review B},
  volume={58},
  number={7},
  pages={4019},
  year={1998},
  publisher={APS}
}

@article{archer2011static,
  title={Static and dynamic properties of type-II composite fermion Wigner crystals},
  author={Archer, Alexander C and Jain, Jainendra K},
  journal={Physical Review B—Condensed Matter and Materials Physics},
  volume={84},
  number={11},
  pages={115139},
  year={2011},
  publisher={APS}
}

@article{chang2005microscopic,
  title={Microscopic Verification of Topological Electron-Vortex Binding in the Lowest Landau-Level Crystal State},
  author={Chang, Chia-Chen and Jeon, Gun Sang and Jain, Jainendra K},
  journal={Physical review letters},
  volume={94},
  number={1},
  pages={016809},
  year={2005},
  publisher={APS}
}

@article{cassella2023discovering,
  title={Discovering quantum phase transitions with fermionic neural networks},
  author={Cassella, Gino and Sutterud, Halvard and Azadi, Sam and Drummond, Neil David and Pfau, David and Spencer, James S and Foulkes, W Matthew C},
  journal={Physical Review Letters},
  volume={130},
  number={3},
  pages={036401},
  year={2023},
  publisher={APS}
}

@article{smith2024unified,
  title={Unified variational approach description of ground-state phases of the two-dimensional electron gas},
  author={Smith, Conor and Chen, Yixiao and Levy, Ryan and Yang, Yubo and Morales, Miguel A and Zhang, Shiwei},
  journal={Physical Review Letters},
  volume={133},
  number={26},
  pages={266504},
  year={2024},
  publisher={APS}
}

@article{pescia2024message,
  title={Message-passing neural quantum states for the homogeneous electron gas},
  author={Pescia, Gabriel and Nys, Jannes and Kim, Jane and Lovato, Alessandro and Carleo, Giuseppe},
  journal={Physical Review B},
  volume={110},
  number={3},
  pages={035108},
  year={2024},
  publisher={APS}
}

@article{luo2023pairing,
  title={Pairing-based graph neural network for simulating quantum materials},
  author={Luo, Di and Dai, David D and Fu, Liang},
  journal={arXiv preprint arXiv:2311.02143},
  year={2023}
}

@article{platzman1993quantum,
  title={Quantum freezing of the fractional quantum Hall liquid},
  author={Platzman, PM and Price, Rodney},
  journal={Physical review letters},
  volume={70},
  number={22},
  pages={3487},
  year={1993},
  publisher={APS}
}

@article{price1993fractional,
  title={Fractional quantum Hall liquid, Wigner solid phase boundary at finite density and magnetic field},
  author={Price, Rodney and Platzman, PM and He, Song},
  journal={Physical review letters},
  volume={70},
  number={3},
  pages={339},
  year={1993},
  publisher={APS}
}

@article{he2005phase,
  title={Phase boundary between the fractional quantum Hall liquid and the Wigner crystal at low filling factors and low temperatures: A path integral Monte Carlo study},
  author={He, WJ and Cui, T and Ma, YM and Chen, CB and Liu, ZM and Zou, GT},
  journal={Physical Review B—Condensed Matter and Materials Physics},
  volume={72},
  number={19},
  pages={195306},
  year={2005},
  publisher={APS}
}

@article{santos1992observation,
  title={Observation of a reentrant insulating phase near the 1/3 fractional quantum Hall liquid in a two-dimensional hole system},
  author={Santos, MB and Suen, YW and Shayegan, M and Li, YP and Engel, LW and Tsui, DC},
  journal={Physical review letters},
  volume={68},
  number={8},
  pages={1188},
  year={1992},
  publisher={APS}
}

@article{santos1992effect,
  title={Effect of Landau-level mixing on quantum-liquid and solid states of two-dimensional hole systems},
  author={Santos, MB and Jo, J and Suen, YW and Engel, LW and Shayegan, M},
  journal={Physical Review B},
  volume={46},
  number={20},
  pages={13639},
  year={1992},
  publisher={APS}
}

@article{pan2005transition,
  title={Transition from a fractional quantum Hall liquid to an electron solid at Landau level filling $\nu$= 1 3 in tilted magnetic fields},
  author={Pan, W and Cs{\'a}thy, GA and Tsui, DC and Pfeiffer, LN and West, KW},
  journal={Physical Review B—Condensed Matter and Materials Physics},
  volume={71},
  number={3},
  pages={035302},
  year={2005},
  publisher={APS}
}

@article{tsui2024direct,
  title={Direct observation of a magnetic-field-induced Wigner crystal},
  author={Tsui, Yen-Chen and He, Minhao and Hu, Yuwen and Lake, Ethan and Wang, Taige and Watanabe, Kenji and Taniguchi, Takashi and Zaletel, Michael P and Yazdani, Ali},
  journal={Nature},
  volume={628},
  number={8007},
  pages={287--292},
  year={2024},
  publisher={Nature Publishing Group UK London}
}

@article{tevsanovic1989hall,
  title={‘‘Hall crystal’’versus Wigner crystal},
  author={Tesanovic, Zlatko and Axel, Francoise and Halperin, BI},
  journal={Physical Review B},
  volume={39},
  number={12},
  pages={8525},
  year={1989},
  publisher={APS}
}

@article{wan2003edge,
  title={Edge reconstruction in the fractional quantum Hall regime},
  author={Wan, Xin and Rezayi, EH and Yang, Kun},
  journal={Physical Review B},
  volume={68},
  number={12},
  pages={125307},
  year={2003},
  publisher={APS}
}

@article{wen1992shift,
  title={Shift and spin vector: New topological quantum numbers for the Hall fluids},
  author={Wen, Xiao-Gang and Zee, A},
  journal={Physical review letters},
  volume={69},
  number={6},
  pages={953},
  year={1992},
  publisher={APS}
}

@article{tsukazaki2010observation,
  title={Observation of the fractional quantum Hall effect in an oxide},
  author={Tsukazaki, A and Akasaka, S and Nakahara, K and Ohno, Y and Ohno, H and Maryenko, D and Ohtomo, A and Kawasaki, M},
  journal={Nature materials},
  volume={9},
  number={11},
  pages={889--893},
  year={2010},
  publisher={Nature Publishing Group UK London}
}

@article{pack2024charge,
  title={Charge-transfer contacts for the measurement of correlated states in high-mobility WSe2},
  author={Pack, Jordan and Guo, Yinjie and Liu, Ziyu and Jessen, Bjarke S and Holtzman, Luke and Liu, Song and Cothrine, Matthew and Watanabe, Kenji and Taniguchi, Takashi and Mandrus, David G and others},
  journal={Nature Nanotechnology},
  volume={19},
  number={7},
  pages={948--954},
  year={2024},
  publisher={Nature Publishing Group UK London}
}

@article{fadon2025extracting,
  title={Extracting Anyon Statistics from Neural Network Fractional Quantum Hall States},
  author={Fadon, Andres Perez and Pfau, David and Spencer, James S and Lou, Wan Tong and Neupert, Titus and Foulkes, WMC},
  journal={arXiv preprint arXiv:2512.15872},
  year={2025}
}

@article{halperin1986compatibility,
  title={Compatibility of crystalline order and the quantized Hall effect},
  author={Halperin, BI and Tesanovic, Z and Axel, F},
  journal={Physical Review Letters},
  volume={57},
  number={7},
  pages={922},
  year={1986},
  publisher={APS}
}

@article{murthy2000hall,
  title={Hall crystal states at $\nu$= 2 and moderate landau level mixing},
  author={Murthy, Ganpathy},
  journal={Physical Review Letters},
  volume={85},
  number={9},
  pages={1954},
  year={2000},
  publisher={APS}
}

@article{kivelson1986cooperative,
  title={Cooperative ring exchange theory of the fractional quantized Hall effect},
  author={Kivelson, Steven and Kallin, C and Arovas, Daniel P and Schrieffer, J Robert},
  journal={Physical review letters},
  volume={56},
  number={8},
  pages={873},
  year={1986},
  publisher={APS}
}

@article{fremling2013coherent,
  title={Coherent state wave functions on a torus with a constant magnetic field},
  author={Fremling, Mikael},
  journal={Journal of Physics A: Mathematical and Theoretical},
  volume={46},
  number={27},
  pages={275302},
  year={2013},
  publisher={IOP Publishing}
}

@article{zhu1993wigner,
  title={Wigner crystallization in the fractional quantum Hall regime: A variational quantum Monte Carlo study},
  author={Zhu, Xuejun and Louie, Steven G},
  journal={Physical review letters},
  volume={70},
  number={3},
  pages={335},
  year={1993},
  publisher={APS}
}

@article{carleo2017solving,
  title={Solving the quantum many-body problem with artificial neural networks},
  author={Carleo, Giuseppe and Troyer, Matthias},
  journal={Science},
  volume={355},
  number={6325},
  pages={602--606},
  year={2017},
  publisher={American Association for the Advancement of Science}
}

@article{vaswani2017attention,
  title={Attention is all you need},
  author={Vaswani, Ashish and Shazeer, Noam and Parmar, Niki and Uszkoreit, Jakob and Jones, Llion and Gomez, Aidan N and Kaiser, Lukasz and Polosukhin, Illia},
  journal={Advances in neural information processing systems},
  volume={30},
  year={2017}
}

@article{qian2025describing,
  title={Describing Landau Level Mixing in Fractional Quantum Hall States with Deep Learning},
  author={Qian, Yubing and Zhao, Tongzhou and Zhang, Jianxiao and Xiang, Tao and Li, Xiang and Chen, Ji},
  journal={Physical Review Letters},
  volume={134},
  number={17},
  pages={176503},
  year={2025},
  publisher={APS}
}

@article{morales2024magic,
  title={Magic angles and fractional Chern insulators in twisted homobilayer transition metal dichalcogenides},
  author={Morales-Dur{\'a}n, Nicol{\'a}s and Wei, Nemin and Shi, Jingtian and MacDonald, Allan H},
  journal={Physical Review Letters},
  volume={132},
  number={9},
  pages={096602},
  year={2024},
  publisher={APS}
}

@article{shi2024adiabatic,
  title={Adiabatic approximation and Aharonov-Casher bands in twisted homobilayer transition metal dichalcogenides},
  author={Shi, Jingtian and Morales-Dur{\'a}n, Nicol{\'a}s and Khalaf, Eslam and MacDonald, Allan H},
  journal={Physical Review B},
  volume={110},
  number={3},
  pages={035130},
  year={2024},
  publisher={APS}
}

@article{paul2023giant,
  title={Giant proximity exchange and flat Chern band in 2D magnet-semiconductor heterostructures},
  author={Paul, Nisarga and Zhang, Yang and Fu, Liang},
  journal={Science Advances},
  volume={9},
  number={8},
  pages={eabn1401},
  year={2023},
  publisher={American Association for the Advancement of Science}
}

@article{wu2019topological,
  title={Topological insulators in twisted transition metal dichalcogenide homobilayers},
  author={Wu, Fengcheng and Lovorn, Timothy and Tutuc, Emanuel and Martin, Ivar and MacDonald, AH},
  journal={Physical review letters},
  volume={122},
  number={8},
  pages={086402},
  year={2019},
  publisher={APS}
}

\onecolumngrid
\newpage
\makeatletter 

\begin{center}
\textbf{\large Supplementary material for: ``\@title ''} \\[10pt]
Ahmed Abouelkomsan$^{1}$ and Liang Fu$^1$ \\
\textit{$^1$Department of Physics, Massachusetts Institute of Technology, Cambridge, MA-02139, USA}\\
\end{center}

\vspace{10pt}

\setcounter{page}{1} 
\setcounter{figure}{0}
\setcounter{section}{0}
\setcounter{equation}{0}
\setcounter{NAT@ctr}{0}

\renewcommand{\thefigure}{S\@arabic\c@figure}
\makeatother

\appendix 

In the supplementary material, we provide additional details about magnetic boundary conditions, the neural network architecture, optimization and a table of the variational energies we obtain. 

\section{Magnetic Boundary Conditions}

Consider a quantum system subject to a  magnetic field
\begin{equation}
B(\boldsymbol r)=B_0+\delta B(\boldsymbol r),
\end{equation}
where \(B_0\) is the uniform (spatially averaged) component and \(\delta B(\boldsymbol r)\) fluctuates in space but averages to zero over the system. Introducing a vector potential via \(B(\boldsymbol r)=\boldsymbol\nabla\times \boldsymbol A(\boldsymbol r)\), it is natural to decompose
\begin{equation}
\boldsymbol A(\boldsymbol r)=\boldsymbol A_0(\boldsymbol r)+\delta \boldsymbol A(\boldsymbol r),\qquad \boldsymbol\nabla\times \boldsymbol A_0 = B_0,
\end{equation}
where \(\boldsymbol A_0(\boldsymbol r)\) is linear in \(\boldsymbol r\) in any gauge. 

We now impose periodic boundary conditions by placing the system on a torus spanned by two basis vectors \(\boldsymbol L_1\) and \(\boldsymbol L_2\). Let \(\boldsymbol L=m\boldsymbol L_1+n\boldsymbol L_2\) be a generic lattice vector with integers \(m,n\). In the presence of a net magnetic flux through the torus, the vector potential $\A(\r)$ cannot be strictly periodic under translations by \(\boldsymbol L\) due to the linear in $\r$ part $\A_0(\r)$,
\begin{equation}
\boldsymbol A(\boldsymbol r+\boldsymbol L)\neq \boldsymbol A(\boldsymbol r),
\end{equation}
On the other hand physical observables must remain invariant under \(\boldsymbol r\mapsto \boldsymbol r+\boldsymbol L\). This invariance is therefore enforced by allowing a translation around the torus to be accompanied by a position-dependent phase of the wavefunction,
\begin{equation}
\psi(\boldsymbol r+\boldsymbol L)=e^{i\xi_{\boldsymbol L}(\boldsymbol r)}\,\psi(\boldsymbol r).
\label{eq:mag_bc}
\end{equation}
Assuming the Hamiltonian depends on \(\boldsymbol A\) only through minimal coupling $
\boldsymbol p\rightarrow \boldsymbol\pi=-i\hbar\boldsymbol\nabla+e\boldsymbol A(\boldsymbol r)
$,  gauge covariance of the kinetic term implies
\(\boldsymbol\pi(\boldsymbol r+\boldsymbol L)\,e^{i\xi_{\boldsymbol L}(\boldsymbol r)}
= e^{i\xi_{\boldsymbol L}(\boldsymbol r)}\,\boldsymbol\pi(\boldsymbol r)\),
so that \(\xi_{\boldsymbol L}(\boldsymbol r)\) compensates the change of \(\boldsymbol A\) under translation. In other words, a torus translation must be accompanied by a gauge transformation so that the Hamiltonian remains gauge-covariant; this requirement determines \(\xi_{\boldsymbol L}(\boldsymbol r)\), 
\begin{equation}
\boldsymbol\nabla \xi_{\boldsymbol L}(\boldsymbol r)=\frac{e}{\hbar}\Big[\boldsymbol A(\boldsymbol r+\boldsymbol L)-\boldsymbol A(\boldsymbol r)\Big].
\label{eq:xi_grad}
\end{equation}

The boundary phases \(\xi_{\boldsymbol L}(\boldsymbol r)\) cannot be chosen independently: they must be mutually consistent when we traverse the torus along different paths. In particular, translating by \(\boldsymbol L_1\) and then \(\boldsymbol L_2\) must bring the wavefunction to the same physical state as translating by \(\boldsymbol L_2\) and then \(\boldsymbol L_1\). Using
\(\psi(\boldsymbol r+\boldsymbol L)=e^{i\xi_{\boldsymbol L}(\boldsymbol r)}\psi(\boldsymbol r)\),
we obtain,
\begin{align}
\psi(\boldsymbol r+\boldsymbol L_1+\boldsymbol L_2)
&=e^{i\xi_{\boldsymbol L_2}(\boldsymbol r+\boldsymbol L_1)}\,e^{i\xi_{\boldsymbol L_1}(\boldsymbol r)}\,\psi(\boldsymbol r),\\
\psi(\boldsymbol r+\boldsymbol L_2+\boldsymbol L_1)
&=e^{i\xi_{\boldsymbol L_1}(\boldsymbol r+\boldsymbol L_2)}\,e^{i\xi_{\boldsymbol L_2}(\boldsymbol r)}\,\psi(\boldsymbol r).
\end{align}
Consistency therefore requires the ratio of these two phases to be unity,
\begin{equation}
\exp\Big\{ i\big[\xi_{\boldsymbol L_2}(\boldsymbol r+\boldsymbol L_1)+\xi_{\boldsymbol L_1}(\boldsymbol r)
-\xi_{\boldsymbol L_1}(\boldsymbol r+\boldsymbol L_2)-\xi_{\boldsymbol L_2}(\boldsymbol r)\big]\Big\}=1.
\label{eq:flux_consistency_phase}
\end{equation}
Taking a gradient and using equation \eqref{eq:xi_grad},
we find,
\begin{equation}
\xi_{\boldsymbol L_2}(\boldsymbol r+\boldsymbol L_1)+\xi_{\boldsymbol L_1}(\boldsymbol r)
-\xi_{\boldsymbol L_1}(\boldsymbol r+\boldsymbol L_2)-\xi_{\boldsymbol L_2}(\boldsymbol r)
=\frac{e}{\hbar}\oint_{\partial\Box}\boldsymbol A\cdot d\boldsymbol\ell
=\frac{e}{\hbar}\int_{\Box} B(\boldsymbol r)\,d^2r
\equiv \frac{e}{\hbar}\Phi.
\end{equation}
Where $\Box$ denotes the torus and $\partial \Box$ denotes the boundary of the torus. Equation \eqref{eq:flux_consistency_phase} then becomes \(\exp\!\big(i\frac{e}{\hbar}\Phi\big)=1\), which implies
\begin{equation}
\Phi = N_\phi \frac{h}{e},\qquad N_\phi\in\mathbb Z,
\end{equation}
i.e., the total magnetic flux through the torus must be an integer multiple of the flux quantum.

 Moreover, the magnetic boundary condition (Eq. \eqref{eq:mag_bc} implies that transporting a particle around the torus returns the wavefunction to itself only up to a phase. The net phase accumulated around the boundary of the torus \(\Box\) is
\begin{equation}
\Delta\varphi \;=\; \frac{e}{\hbar}\oint_{\partial\Box}\boldsymbol A\cdot d\boldsymbol\ell
\;=\;\frac{e}{\hbar}\Phi \;=\; 2\pi N_\phi,
\end{equation}
 The single particle wavefunction therefore carries an integer total vorticity set by the number of flux quanta $N_\phi$. 

For the wavefunction to be continuous, the net winding necessarily implies the existence of zeros of in the fundamental domain of the torus. The winding around a  zero can be either positive (vortex) or negative (anti-vortex) but the total winding is equal to $N_\phi$. Let $N_{z}$ be the number of vortices and $N_{\bar{z}}$ be the number of anti-vortices with negative winding. Assuming $B_0 > 0$, we have \begin{equation}
    N_{\phi} = N_{z} - N_{\bar{z}}
\end{equation}

Around a vortex, a complex function $f(z,\bar{z}) \sim z^n$ for some integer $n$ Similariy, for an anti-vortex, $f(z,\bar{z}) \sim \bar{z}^n$. Taking case of the lowest landau level (LLL) where the wavefunction is holormorphic (up to a Gaussian factor), we immediately see that any LLL wavefunction has $N_\phi$ zeros. 


In addition to position-dependent magnetic phases $\xi_{\L}(\r)$ due to the existence of net flux, the wavefunction can acquire a twist phase $e^{i \varphi}$ under translations, corresponding to threading a flux through the handles of the torus which is implemented by a \textit{constant} vector potential.

The above discussions generalizes readily for a many-body system. In this case, the  wavefunction of $N$ particles transforms as, 

\begin{equation}
\label{eq:magBC_supp}
    \Psi(\r_1, \cdots, \r_i + \L, \cdots, \r_N) = e^{i\varphi} e^{i \xi_{\L}(\r_i)}  \Psi(\r_1, \cdots, \r_N)
\end{equation}
under torus translations of any particle $\r_i$. 


\section{Neural Network Architecture}

In this section, we describe the MagNet architecture used throughout the paper. We use an architecture that is schematically shown in Fig. 1 in the main text and has both self-attention and multilayer perceptron components. As we are working with a torus supercell, we use periodic features for each particle. First a feature vector $f_i \equiv f(\r_i)$ is defined for the $i$-th particle as \begin{equation}
    \f_i \equiv f(\r_i) = \begin{pmatrix}
        \cos(\G_1 \cdot \r_i) \\
        \cos(\G_2 \cdot \r_i) \\
        \sin(\G_1 \cdot \r_i) \\
        \sin(\G_2 \cdot \r_i)
    \end{pmatrix}
\end{equation}
 where $\G_{a}$ ($ a= 1,2$) denote the two supercell reciprocal vectors $\G_a \cdot \L_b = 2 \pi \delta_{ab}$ with $\L_a$ ($a = 1,2$) are the two basis vector of the torus. 
This feature vector represent a periodized version of the particle coordinates. Each feature vector is then mapped to an initial $\h_i^{0}$ vector which lives in a higher dimensional space through a transformation, \begin{equation}
    \h_i^{0} = W^{0} \f_i 
\end{equation}
with $W^0 \in \mathbb{R}^{d_L} \times \mathbb{R}^{2 d_{\rm dim}}$ with $d_L$ is the dimension of the higher dimensional space where the particle coordinates are embedded which corresponds to the width of the internal layers of the neural network. $d_{\rm dim}$ is the spatial dimension ($d_{\rm dim} = 2$ in our case). Importantly, $W^0$ is the same for each particle coordinate. Having mapped the coordinates of all particles to vectors $\{\h_i^0\}$, these vectors undergo two different types of transformations, multi-head self attention followed by multilayer perceptron (MLP) which we explain next. 
\subsection{Multilayer Perceptron (MLP)}
MLPs are standard feed-forward neural networks that implements the following transformation on a \textit{generic} vector $\g_1 \in \mathbb{R}^{d_L}$ to yield another vector $\g_2$, \begin{equation}
\label{eq:MLP}
    \g_2 = \g_1 + F(W \g_1 + \b)
\end{equation}
with $W$ is a linear transformation $W \in \mathbb{R}^{d_L} \times \mathbb{R}^{d_L}$ and $\b \in \mathbb{R}^{d_L}$ is a bias vector. $F$ here represents a non-linear activation function which we choose to be the GELU function. 
\subsection{Self-attention}
In the NN architecture (Fig. 1 in the main text), MLPs act individually on each particle without mixing different particle streams which cannot describe correlations between particles. In order to capture such correlations, we utilize the self-attention mechanism \cite{vaswani2017attention} which form the basis of transformers used in large language models. Self-attention mechanism learns how each particle is influenced by the remaining particles. Self attention acts of all set of particle streams in the $l$-layer $\{\h_i^{l}\}$ to generate intermediate streams $\{\g_i^{l}\}$ that are then passed to the MLP \eqref{eq:MLP}. Schematically we have the following, 
\begin{equation}
    \cdots \rightarrow \{\h_i^{l}\} \xrightarrow{\rm SELF-ATTN}  \{\g_i^{l}\} \xrightarrow{\rm PERCEPTRON} \{\h_i^{l+1}\} \rightarrow \cdots 
\end{equation}

In the self-attention mechanism, three features for each element in $\{h_i^{l}\}$ are defined which are called \textit{keys}, \textit{queries} and \textit{values} and are defined as,
\begin{equation}
    \k_i^{l h} = W_{k}^{lh} \h_i^{l}, \> \> W_{k}^{lh} \in \mathbb{R}^{d_{\rm attn}} \times \mathbb{R}^{d_L}
\end{equation}
\begin{equation}
    \q_i^{l h} = W_{q}^{lh} \h_i^{l}, \> \> W_{q}^{lh} \in \mathbb{R}^{d_{\rm attn}} \times \mathbb{R}^{d_L}
\end{equation}
\begin{equation}
    \v_i^{l h} = W_{v}^{lh} \h_i^{l}, \> \> W_{v}^{lh} \in \mathbb{R}^{d_{\rm attnvals}} \times \mathbb{R}^{d_L}
\end{equation}
The transformations $\{W_{k}^{lh},W_{q}^{lh}, W_{v}^{lh} \} $ are the same for each particle $i$. We use multi-head attention where more than one self-attention operation is applied, indexed by $h$ in $W_{k}^{lh}$ , $W_{q}^{lh}$ and $ W_{v}^{lh}$ which is not to be confused with the particle stream vectors $\h_i^{l}$. The dimensions $d_{\rm attn}$ and $d_{\rm attnvals}$ are different from $d_{L}$ and are generally much smaller.  The keys and queries vectors have the same vector space dimension $d_{\rm attn}$ to allow for comparison between two different particle streams $i,j$ through the dot product $\k_i^{lh} \cdot \q_j^{lh}$. The values $\v_i^{lh}$ is a measure of the influence on particle $i$ from the remaining particles $j \neq i$. The value of self-attention for each particle $i$ is computed as \begin{equation}
\label{eq:selfattn}
    {\rm SELFATTN}_i(\{\h_j^l\}; W_{k}^{lh},W_{q}^{lh}, W_{v}^{lh}) = \frac{1}{\mathcal{N}} \sum_{j = 1}^N {\rm exp}(\k_i^{lh} \cdot \q_j^{lh}) \v_j^{lh} 
\end{equation}
with a normalization factor, \begin{equation}
    \mathcal{N} = \sqrt{d_{\rm attnvals}} \sum_{j = 1}^N {\rm exp}(\k_i^{lh} \cdot \q_j^{lh})
\end{equation}
The physical meaning of \eqref{eq:selfattn} is transparent. The self-attention value of particle $i$ returns the value $\v_j^{l h}$ of particle stream $j$ weighted by the exponential factor which measures the correlations between $i,j$ such that the most relevant particles $j$ for each particle $j$ are identified. The intermediate streams are given by \begin{equation}
    \g_i^{l+1} = \h_i^{l} + W_o^{l} \> {\rm concat}_h \> [{\rm SELFATTN}_i(\{\h_j^l\}; W_{k}^{lh},W_{q}^{lh}, W_{v}^{lh})]
\end{equation}
All values of self-attention from the multi-head attention are concatenated then transformed through transformation $W_0^{l} \in \mathbb{R}^{d_L} \times \mathbb{R}^{N_{\rm heads} \times {\rm attnvals}}$. The intermediate streams $\{\g_i^{l}\}$ are passed through MLP giving rise to,
\begin{equation}
    \h_i^{l+1} = \g_i^{l} + F(W^{l} \g_i^{l} + \b^{l+1})
\end{equation}
Here $W^l$ and $\b^{l}$ are the same for all particle streams which are labeled by $i$. 
\subsection{Mapping onto $N \times N_{\rm det}$ amplitudes}
The final output of the neural network $\{\h_i^L\}$ is mapped onto $N \times N_{\rm det} $ distinct sets of amplitudes
 \begin{equation}
    F_j^n(\r_i; \{\r_{\neq i }\}) = \w_{2 j}^n \cdot \h_i^{L} + i \w_{2 j+1}^n \cdot \h_i^{L}
\end{equation}
with $j = 1 , \cdots,  N$ and $n = 1, \cdots, N_{\rm det}$.  $\w_{2j}^n$ and $\w_{2j+1}^n$ are matrices that construct the real and the imaginary part of $F_j^n$. 
It is important to note that in order for the full ansatz  to describe a fermionic anti-symmetric wavefunction, the amplitudes $F_j^n(\r_i; \{\r_{\neq i }\})$ are chosen to be permutation invariant in the coordinates $\{\r_{\neq i }\}$,
\begin{equation}
    F_j^n(\r_i; \{\r_k, \cdots, \r_l,\cdots\}) =  F_j^n(\r_i; \{\r_l, \cdots, \r_k,\cdots\})
\end{equation}
for $k,l \neq i$. This property is inherited from the form of the self-attention operations \eqref{eq:selfattn}. 
\subsection{Mapping onto $ N_\phi \times N \times N_{\rm det}$ winding maps }

In addition to the set of amplitudes $F_j^n$, the final output of the neural network is also mapped to construct the winding maps $\eta_j^{(n,\alpha)}$ as symmetric many-body functions,
\begin{equation}
    \eta_j^{(n,\alpha)}(\{\r\}) = \dfrac{1}{N} \sum_{i = 1, \cdots N} \boldsymbol{\Omega}_{2 j}^{(n,\alpha)} \cdot \h_i^{L} + i \boldsymbol{\Omega}_{2 j+1}^{(n,\alpha)}  \cdot \h_i^{L}
\end{equation}
with $i = 1 , \cdots,  N$ , $n = 1, \cdots, N_{\rm det}$ and $\alpha = 1, \cdots, N_\phi$. $\boldsymbol{\Omega}_{2 j}^{(n,\alpha)}$ and $\boldsymbol{\Omega}_{2 j+1}^{(n,\alpha)}$ are two matrices that construct the real and imaginary part of the complex coordinate of the zero. 

The sum of the winding maps,
\(\overline{\eta}^{(n)}_j=\big(\sum_{\alpha}\eta^{(n,\alpha)}_j\big)\bmod(L_x+iL_y)\),
fixes the twist phase $\varphi$ in Eq.~\eqref{eq:magBC_supp} \cite{wang2019lattice}. We choose a twist phase of $\varphi =0$ by imposing that $\eta_j^{(n)} = 0$ by first obtaining  $\eta_j^{(n,\alpha)}(\{\r\})$ and subtracting it from its mean over $\alpha$. 

\begin{table}
    \centering
    \renewcommand{\arraystretch}{1.3}
    \begin{tabular}{ll l}
        \toprule
        \textbf{Parameter} & & \textbf{Value} \\
        \midrule
        \textbf{Architecture} & & \\ 
        Network layers & & $L = 4$ \\
        Attention heads per layer & & $N_{\rm heads} = 4$ \\
        Attention dimension & & $d_{\rm attn} = d_{\rm attnvals} = 64$ \\
        Perceptron dimension & & $d_L = 256$ \\
        $\#$ perceptrons per layer & & $2$ \\
         Determinants & & $N_{\rm det} = 2$\\
         Layer norm & & True \\
         Activation & & GELU \\
        \midrule
        \textbf{Training} & & \\ 
        Training iterations & & $15e6 - 20e6$ \\
        Learning rate (fixed) &  & $\eta_0 = 0.1$  \\
        Optimizer &  & KFAC \\
        \midrule
        \textbf{MCMC} & & \\
        Batch size & & $1024$ \\
        Initial move width & & $0.2$ \\
        \midrule
        \textbf{KFAC} & & \\ 
        Norm constraint & & $1 \times 10^{-4}$ \\
        Damping & & $1 \times 10^{-3}$ \\
        $L_2$ regularization & & 0.0 \\
        Momentum & & 0.0 \\
        \bottomrule
    \end{tabular}
    \caption{Table of default hyperparameters used in our numerical calculations with the MagNet architecture.}
    \label{Tab:Hyperparams}
\end{table}

\subsection{Jastrow factor}
\textcolor{black}{To enforce the Coulomb cusp condition, we use a simple Jastrow factor that takes the form
\begin{equation}
    \mathcal{J}(\r_1 , \cdots ,\r_N) = \sum_{i} \sum_{j \neq i} u(|\r_i - \r_j|)  
\end{equation}
where \begin{equation}
    u(r) = \frac{-1}{3} \dfrac{\alpha^2}{\alpha + \r}
\end{equation}
with $\alpha$ being a single trainable parameter. }

\section{Optimization}

In Table \ref{Tab:Hyperparams}, we list the hyperparameters of the neural network we used in our work. We particularly note the small norm constraint we use in order to stabilize the training.
We initialize the training in two different ways. For $\kappa  = 3$ and $\kappa = 9$, we do a random initialization while for $\kappa = 12, 15, 20$, we initialize from the optimized phase at either $\kappa = 3$ or $\kappa = 9$. We find that initializing from the optimized FQH phase to accelerate convergence, even when the ground state exhibit crystalline correlations. We checked in various cases that both random initialization and initialization from the optimized FQH liquid give the same result.

\begin{figure}[h!]
    \centering
    \includegraphics[width=0.7\linewidth]{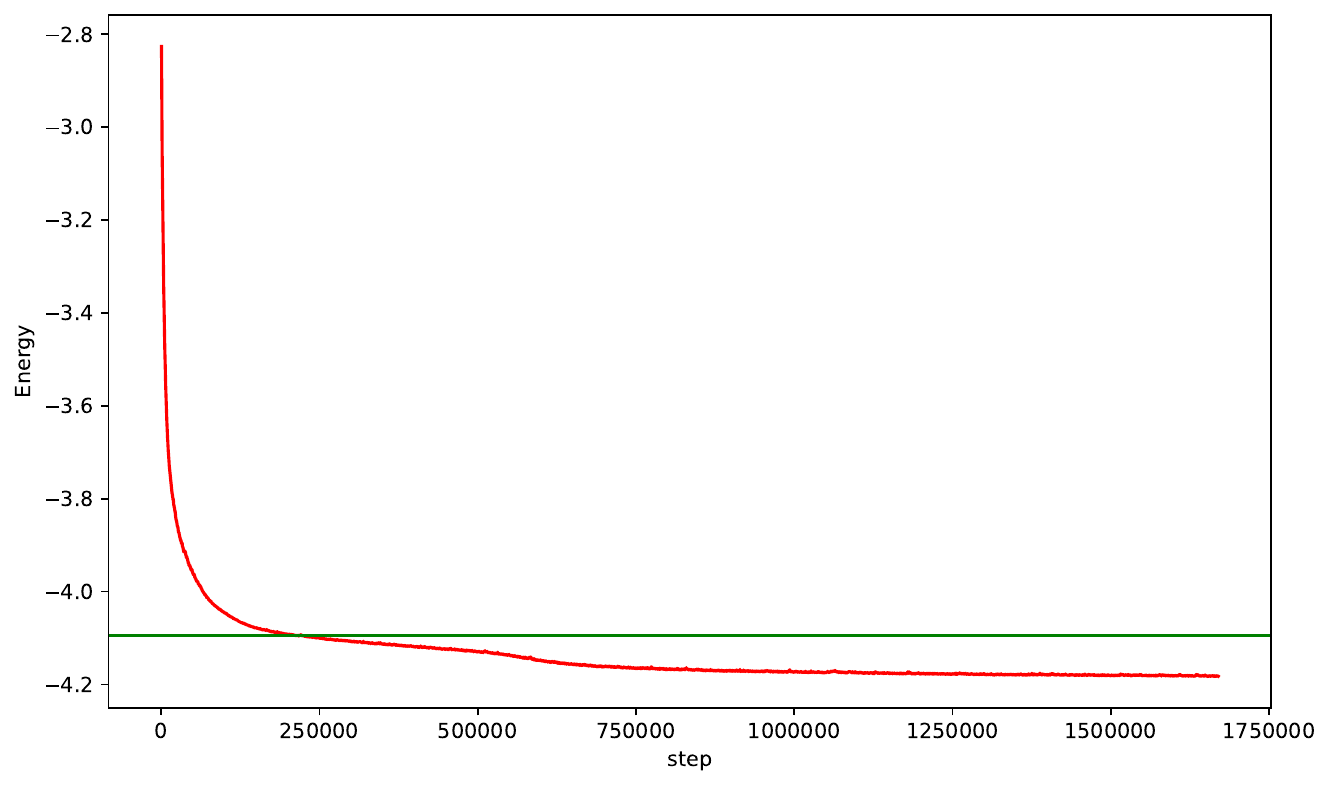}
    \caption{Training curve for $N = 12$ and $N = 36$ at $\kappa = 3.0$ with random initialization. Energy at a given step is averaged over the previous 4000 steps and it is in units of $\hbar \omega_c$ and does not include the Madelung constant of this finite torus. The green line denotes the energy of the lowest landau level projected ED.  }
    \label{fig:traningcurve}
\end{figure}

Fig. \ref{fig:traningcurve} shows a representative training curve for $N= 12 $ and $N_\phi = 36$ flux quantum at $\kappa = 3.0$

\section{Table of Energies}
Below, we list the energies we obtained for various Landau level mixing $\kappa$ for both $ N = 9 $ and $ N = 12$. We measure energies in units of $\hbar \omega_c$. To facilitate comparison with exact diagonalization, we do not include the Madelung constant in the table below. 
\begin{table}[h!]
  \centering
  \begin{tabular}{|l|c|c|}
    \hline
     Landau level mixing & $N = 9$ & $N = 12$ \\
    \hline
    $\kappa = 3$ & -2.61141(27) & -4.18057(20) \\
    \hline
    $\kappa = 9$ & -17.24265(11) & -25.06976(21) \\
    \hline
    $\kappa = 12$ & -24.680162(89) & -35.66683(21) \\
    \hline
    $\kappa = 15$ & -32.167333(99) & -46.35916(16) \\
    \hline
    $\kappa = 20$ & -44.753212(87) & -64.29958(12) \\
    \hline
  \end{tabular}
  \end{table}
\end{document}